\documentclass[fleqn,12pt,twoside]{article}
\usepackage{espcrc1}

\usepackage{graphicx}
\title{Energy Dependence of $\Lambda$ and $\bar{\Lambda}$ Production at 
       CERN-SPS Energies}

\author{A. Mischke$^{8}$
        for the NA49 Collaboration}

\begin{document}
\maketitle

%%%%%%%%%%%%%%%%%%%%%%%%%%%%%%%%%%%%%%%%%%%%%%%%%%%%%%%%%%%%%%%%%%%%%%%%%%%%%%
%                            author list
%%%%%%%%%%%%%%%%%%%%%%%%%%%%%%%%%%%%%%%%%%%%%%%%%%%%%%%%%%%%%%%%%%%%%%%%%%%%%%
\noindent

S.V.~Afanasiev$^{9}$,T.~Anticic$^{21}$, B.~Baatar$^{9}$,D.~Barna$^{5}$,
J.~Bartke$^{7}$, R.A.~Barton$^{3}$, M.~Behler$^{15}$,
L.~Betev$^{10}$, H.~Bia{\l}\-kowska$^{19}$, A.~Billmeier$^{10}$,
C.~Blume$^{8}$, C.O.~Blyth$^{3}$, B.~Boimska$^{19}$, M.~Botje$^{1}$,
J.~Bracinik$^{4}$, R.~Bramm$^{10}$, R.~Brun$^{11}$,
P.~Bun\v{c}i\'{c}$^{10,11}$, V.~Cerny$^{4}$, O.~Chvala$^{17}$,
J.G.~Cramer$^{18}$, P.~Csat\'{o}$^{5}$, P.~Dinkelaker$^{10}$,
V.~Eckardt$^{16}$, P.~Filip$^{16}$,
H.G.~Fischer$^{11}$, Z.~Fodor$^{5}$, P.~Foka$^{8}$, P.~Freund$^{16}$,
V.~Friese$^{8,15}$, J.~G\'{a}l$^{5}$,
M.~Ga\'zdzicki$^{10}$, G.~Georgopoulos$^{2}$, E.~G{\l}adysz$^{7}$,
S.~Hegyi$^{5}$, C.~H\"{o}hne$^{15}$, G.~Igo$^{14}$,
P.G.~Jones$^{3}$, K.~Kadija$^{11,21}$, A.~Karev$^{16}$,
V.I.~Kolesnikov$^{9}$, T.~Kollegger$^{10}$, M.~Kowalski$^{7}$,
I.~Kraus$^{8}$, M.~Kreps$^{4}$, M.~van~Leeuwen$^{1}$,
P.~L\'{e}vai$^{5}$, A.I.~Malakhov$^{9}$, S.~Margetis$^{13}$,
C.~Markert$^{8}$, B.W.~Mayes$^{12}$, G.L.~Melkumov$^{9}$,
C.~Meurer$^{10}$,
A.~Mischke$^{8}$, M.~Mitrovski$^{10}$,
J.~Moln\'{a}r$^{5}$, J.M.~Nelson$^{3}$,
G.~P\'{a}lla$^{5}$, A.D.~Panagiotou$^{2}$,
K.~Perl$^{20}$, A.~Petridis$^{2}$, M.~Pikna$^{4}$, L.~Pinsky$^{12}$,
F.~P\"{u}hlhofer$^{15}$,
J.G.~Reid$^{18}$, R.~Renfordt$^{10}$, W.~Retyk$^{20}$,
C.~Roland$^{6}$, G.~Roland$^{6}$, A.~Rybicki$^{7}$, T.~Sammer$^{16}$,
A.~Sandoval$^{8}$, H.~Sann$^{8}$, N.~Schmitz$^{16}$, P.~Seyboth$^{16}$,
F.~Sikl\'{e}r$^{5}$, B.~Sitar$^{4}$, E.~Skrzypczak$^{20}$,
G.T.A.~Squier$^{3}$, R.~Stock$^{10}$, H.~Str\"{o}bele$^{10}$, T.~Susa$^{21}$,
I.~Szentp\'{e}tery$^{5}$, J.~Sziklai$^{5}$,
T.A.~Trainor$^{18}$, D.~Varga$^{5}$, M.~Vassiliou$^{2}$,
G.I.~Veres$^{5}$, G.~Vesztergombi$^{5}$,
D.~Vrani\'{c}$^{8}$, S.~Wenig$^{11}$, A.~Wetzler$^{10}$, C.~Whitten$^{14}$,
I.K.~Yoo$^{8,15}$, J.~Zaranek$^{10}$, J.~Zim\'{a}nyi$^{5}$       

\vspace{0.3cm}
\begin{footnotesize}
\noindent
$^{1}$NIKHEF, Amsterdam, Netherlands. \\
$^{2}$Department of Physics, University of Athens, Athens, Greece.\\
$^{3}$Birmingham University, Birmingham, England.\\
$^{4}$Comenius University, Bratislava, Slovakia.\\
$^{5}$KFKI Research Institute for Particle and Nuclear Physics, Budapest, Hungary.\\
$^{6}$MIT, Cambridge, USA.\\
$^{7}$Institute of Nuclear Physics, Cracow, Poland.\\
$^{8}$Gesellschaft f\"{u}r Schwerionenforschung (GSI), Darmstadt, Germany.\\
$^{9}$Joint Institute for Nuclear Research, Dubna, Russia.\\
$^{10}$Fachbereich Physik der Universit\"{a}t, Frankfurt, Germany.\\
$^{11}$CERN, Geneva, Switzerland.\\
$^{12}$University of Houston, Houston, TX, USA.\\
$^{13}$Kent State University, Kent, OH, USA.\\
$^{14}$University of California at Los Angeles, Los Angeles, USA.\\
$^{15}$Fachbereich Physik der Universit\"{a}t, Marburg, Germany.\\
$^{16}$Max-Planck-Institut f\"{u}r Physik, Munich, Germany.\\
$^{17}$Institute of Particle and Nuclear Physics, Charles University, Prague, Czech Republic.\\
$^{18}$Nuclear Physics Laboratory, University of Washington, Seattle, WA, USA.\\
$^{19}$Institute for Nuclear Studies, Warsaw, Poland.\\
$^{20}$Institute for Experimental Physics, University of Warsaw, Warsaw, Poland.\\
$^{21}$Rudjer Boskovic Institute, Zagreb, Croatia.\\
\end{footnotesize}

%==============================================================================
\begin{abstract}
%\begin{footnotesize}
Rapidity distributions for $\Lambda$ and $\bar{\Lambda}$ hyperons in 
central Pb-Pb collisions at 40, 80 and 158~A$\cdot$GeV and for 
${\rm K}_{s}^{0}$ mesons at 158~A$\cdot$GeV are presented.
The lambda multiplicities are studied as a function of collision
energy together with AGS and RHIC measurements and compared to model
predictions. A different energy dependence of the $\Lambda/\pi$ and
$\bar{\Lambda}/\pi$ is observed.   
The $\bar{\Lambda}$/$\Lambda$ ratio shows a steep increase with
collision energy. 
Evidence for a $\bar{\Lambda}/\bar{\rm p}$ ratio greater than 1 is
found at 40~A$\cdot$GeV.
%\end{footnotesize}
\end{abstract}

%==============================================================================
\section{Introduction}
\vspace{-0.2cm}
Anomalies in the energy dependence of strangeness production have been
predicted as a hint for the onset of deconfinement~\cite{Gaz96,GazGor99}.
Since $\Lambda$ hyperons contain between 30 and 60$\%$ of the
total strangeness produced in hadronic interactions
their measurements allows to study simultaneously
strangeness production and the effect of net baryon density.

%==============================================================================
\section{Analysis}
\vspace{-0.2cm}
The data sets used for the present analysis are the 7.2~$\%$, 7.2~$\%$
and 10~$\%$ most central Pb-Pb events at 40, 80 and 158 A$\cdot$GeV
beam energy, corresponding to 8.73, 12.3 and 17.3 GeV c.m. energy per 
nucleon nucleon pair. 
The NA49 experiment~\cite{NIM99} identifies neutral strange baryons
by reconstructing their characteristic V0 decay topology
\mbox{$\Lambda \rightarrow p+\pi^{-}$}, \mbox{$\bar{\Lambda}
\rightarrow \bar{p}+\pi^{+}$} and \mbox{${\rm K}_{s}^{0} \rightarrow
\pi^{+}+\pi^{-}$}. The $\Lambda$ hyperons contain the short-lived
$\Sigma^{0}$, which decay electro-magnetically into $\Lambda\gamma$.

The charged decay products are measured with four time projection
chambers (TPCs), two of them are located inside two large dipole
magnets, the other two downstream of the magnets symmetrically 
to the beam line~\cite{NIM99}. 
\begin{figure}[t]
%\hspace{-0.5cm}
\begin{minipage}[b]{6cm}
 \begin{center}
 \includegraphics[width=6cm,height=7cm]{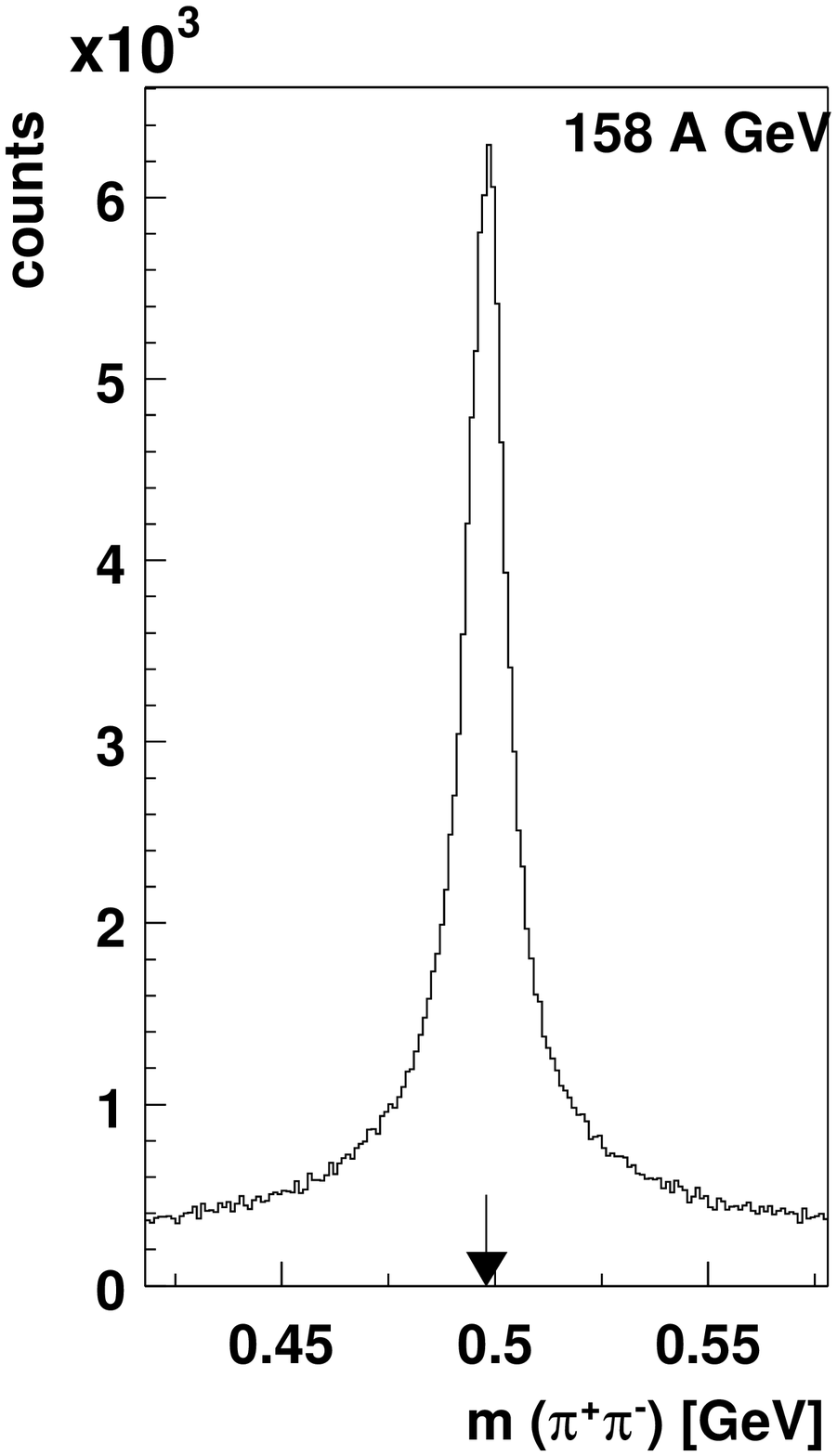}
 \end{center}
\end{minipage}
\hspace{.5cm}
%\hspace{\fill}
%
\begin{minipage}[b]{6cm}
 \begin{center}
 \includegraphics[width=9cm]{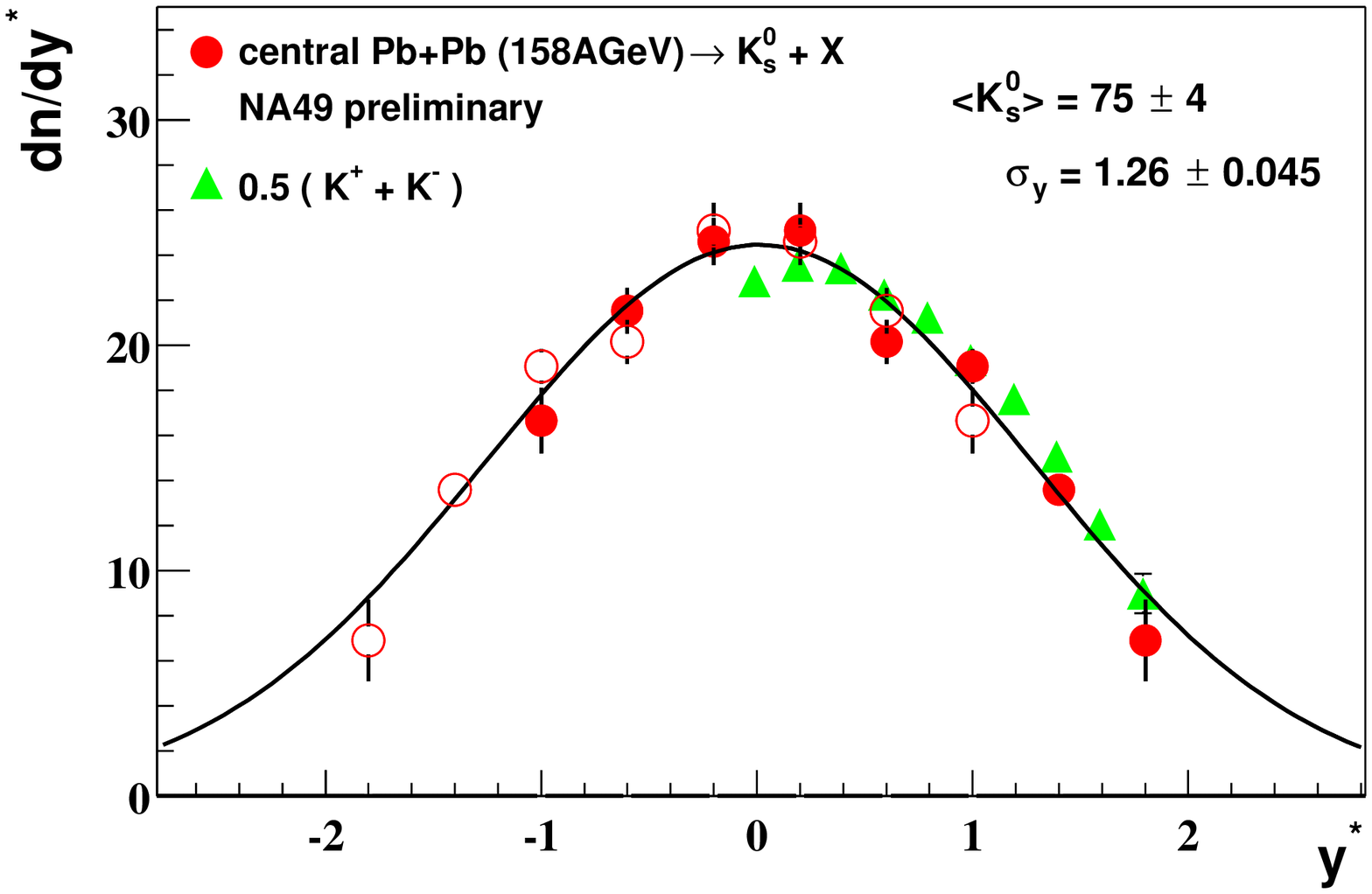}
 \end{center}
 \end{minipage}
 \vspace{-1.cm}
 \caption{\protect \footnotesize Invariant mass distribution (left) and
rapidity distribution (right) of ${\rm K}_{s}^{0}$ at 158~A$\cdot$GeV.}
 \label{fig2}
\end{figure}
In figure~\ref{fig2}, the invariant mass distribution of 
${\rm K}_{s}^{0}$ at 158 A$\cdot$GeV is shown. An agreement
between the peak position and the nominal ${\rm K}_{s}^{0}$ mass, 
indicated by the arrow, is observed. The invariant mass distributions
of $\Lambda$ and $\bar{\Lambda}$ are shown in reference~\cite{Mis02}. 
The mass resolution ($\sigma_{\rm m}$) is 2~MeV/$c^{2}$ for the lambdas
and 4~MeV/$c^{2}$ for ${\rm K}_{s}^{0}$.

%==============================================================================
\section{Spectra}
\vspace{-0.2cm}
\begin{figure}[t]
\hspace{-.9cm}
\begin{minipage}[b]{18cm}
 \begin{center}
 \includegraphics[width=14.5cm,height=6.08cm]{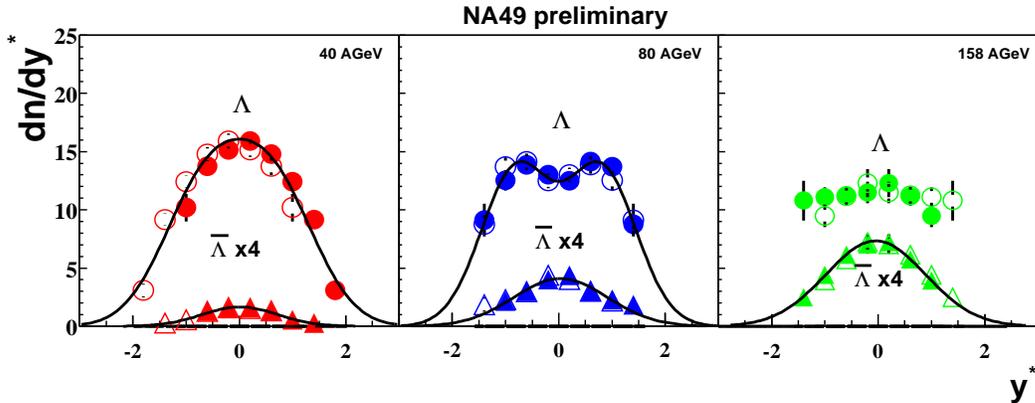}
 \end{center}
\end{minipage}
 \vspace{-2.1cm}
 \caption{\protect \footnotesize Rapidity distribution of $\Lambda$ and
$\bar{\Lambda}$ produced in central Pb-Pb collisions at 40 (left), 80
(middle) and 158~A$\cdot$GeV/u (right). The reflected points
(open symbols) are in good agreement with the measured ones (filled
symbols).}     
 \label{fig1}
\end{figure}
Corrections, described in detail in reference~\cite{Mis02},
are applied bin by bin in rapidity and transverse momentum for
geometrical acceptance and tracking efficiency.
The rapidity distributions are obtained by integrating the measured
$p_{\rm T}$-spectra and by extrapolation into unmeasured regions. 
In figure~\ref{fig1},
the rapidity distributions of $\Lambda$ and $\bar{\Lambda}$ are
summarized for all three energies. 
It is found that the distribution of the $\Lambda$ is
broader than that of the $\bar{\Lambda}$. 
The $\Lambda$ rapidity distribution becomes broader with increasing
energy. 
The total lambda multiplicities are obtained by integration of the
rapidity spectra with small extrapolations into unmeasured
regions using a Gaussian fit for the $\bar{\Lambda}$ at all three
energies and the $\Lambda$ at 40 A$\cdot$GeV. A double Gaussian is
used for the $\Lambda$ at 80 A$\cdot$GeV. For the $\Lambda$ rapidity
distribution at 158~A$\cdot$GeV an extrapolation is made using
realistic estimates of the tails (e.g. $\Lambda$ from central S+S and
net-proton distribution at 158~A$\cdot$GeV~\cite{Mis02}).   

The corrections and the analysis procedure were checked by extracting
the ${\rm K}_{s}^{0}$ meson at 158~A$\cdot$GeV and comparing them to
the charged kaons~\cite{KPI02}. The charged kaons are identified with
a different method (dE/dx). The comparison between the rapidity
distribution of ${\rm K}_{s}^{0}$ and the charged kaons (using
isospin symmetry: ${\rm K}_{s}^{0} = ({\rm K}^{+}+{\rm K}^{-})/2$)  
is shown in figure~\ref{fig2} (right). 
Good agreement is observed. The total ${\rm K}_{s}^{0}$ multiplicity,
quantified using a Gaussian fit to extrapolate the unmeasured regions
yields $\langle{\rm K}_{s}^{0}\rangle$ = 75 $\pm$ 4~. 

%The transverse mass ($m_{\rm T}=\sqrt{p_{\rm T}^2+m_0^2}$, where $m_0$
%is the rest mass of the particle) distributions of $\Lambda$,
%$\bar{\Lambda}$ and ${\rm K}_{s}^{0}$ (at mid-rapidity) are plotted in
%figure~\ref{fig1}.  
%All spectra follow a Boltzmann like function in $m_{\rm T}$ according
%to 1/$m_{\rm T}\cdot {\rm d}^2n/({\rm d}m_{\rm T} {\rm d}y) \propto
%e^{- m_{\rm T}/{T}}$, where $T$ is the inverse slope parameter.  

%==============================================================================
\section{Energy Dependence}
\vspace{-0.2cm}
\begin{figure}[t]
\begin{minipage}[b]{6cm}
 \begin{center}
 \includegraphics[width=8.2cm,height=6cm]{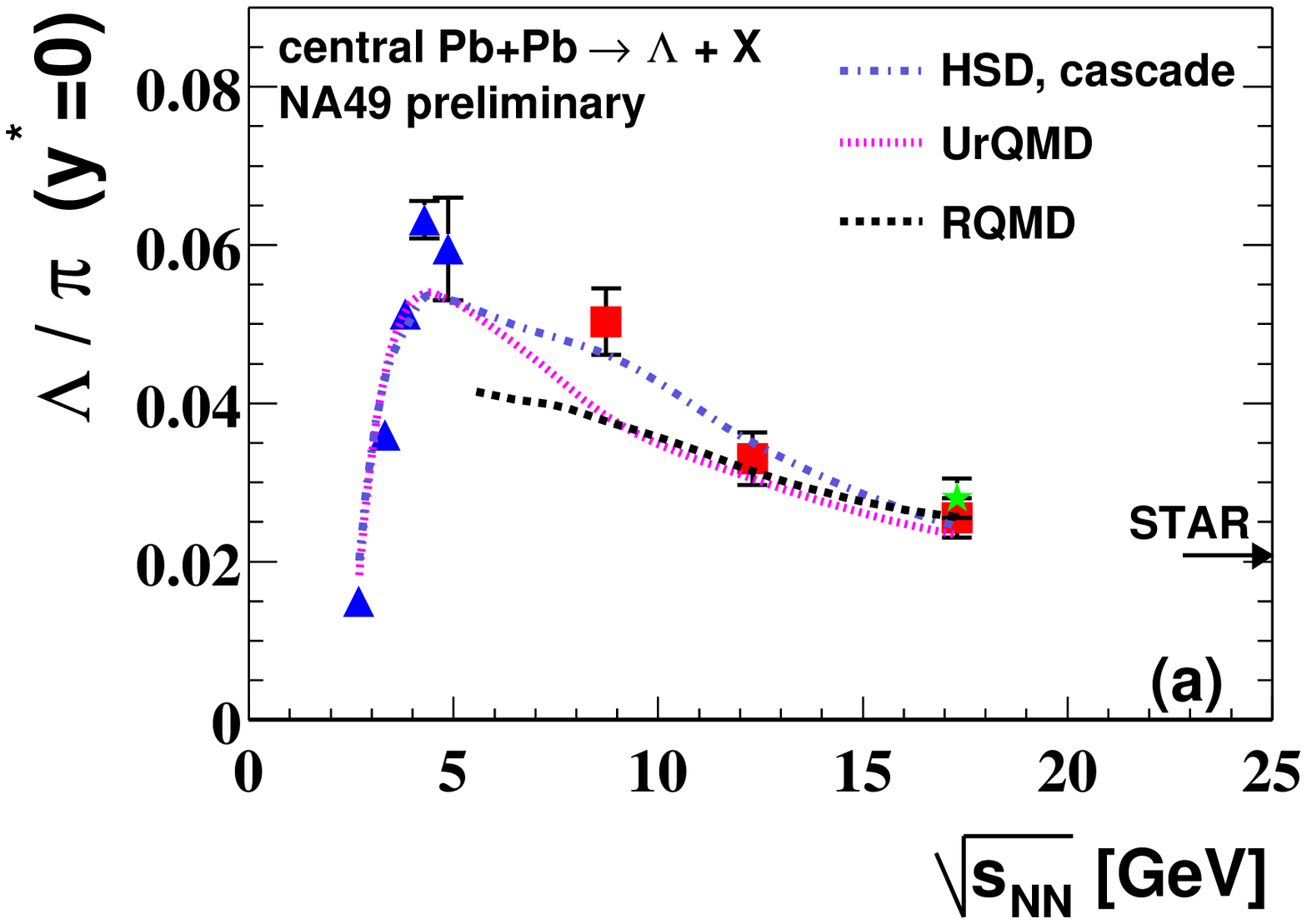}
 \end{center}
\end{minipage}
\hspace{1.5cm}
%\hspace{\fill}
%
\begin{minipage}[b]{6cm}
 \begin{center}
 \includegraphics[width=8.2cm,height=6cm]{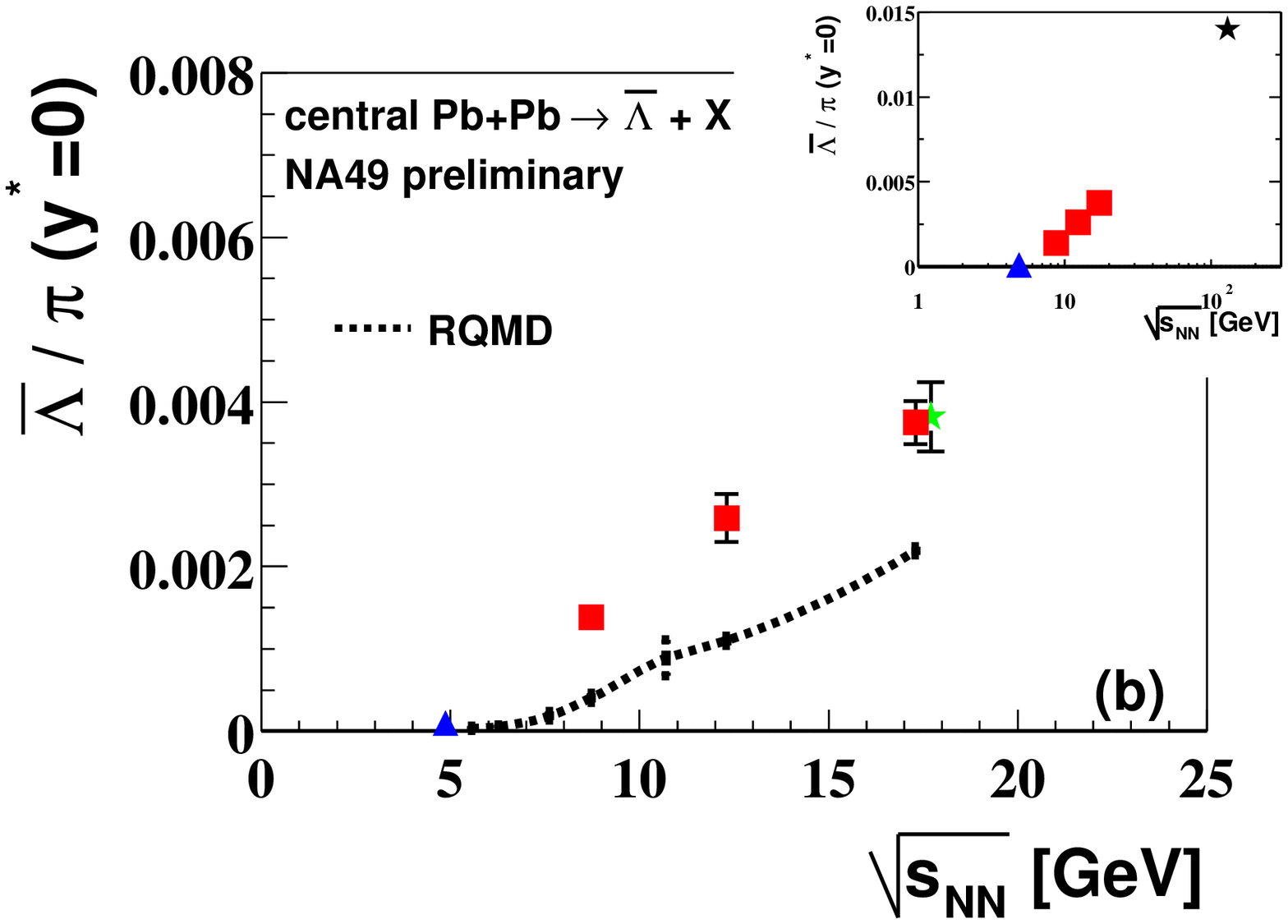}
 \end{center}
\end{minipage}

\vspace{-0.5cm}

\begin{minipage}[b]{6cm}
 \begin{center}
 \includegraphics[width=8.2cm,height=6cm]{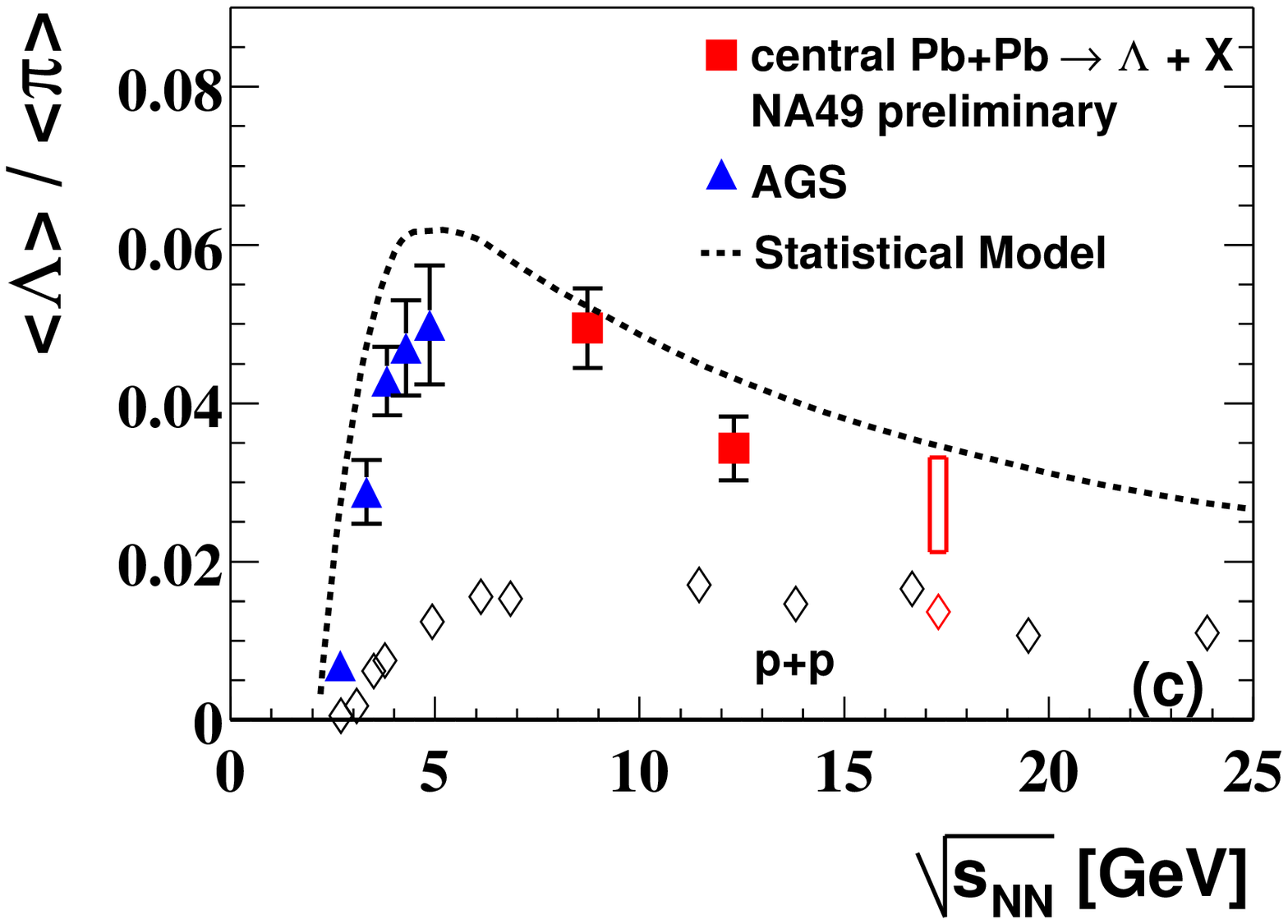}
 \end{center}
\end{minipage}
\hspace{1.5cm}
%\hspace{\fill}
%
\begin{minipage}[b]{6cm}
 \begin{center}
 \includegraphics[width=8.2cm,height=6cm]{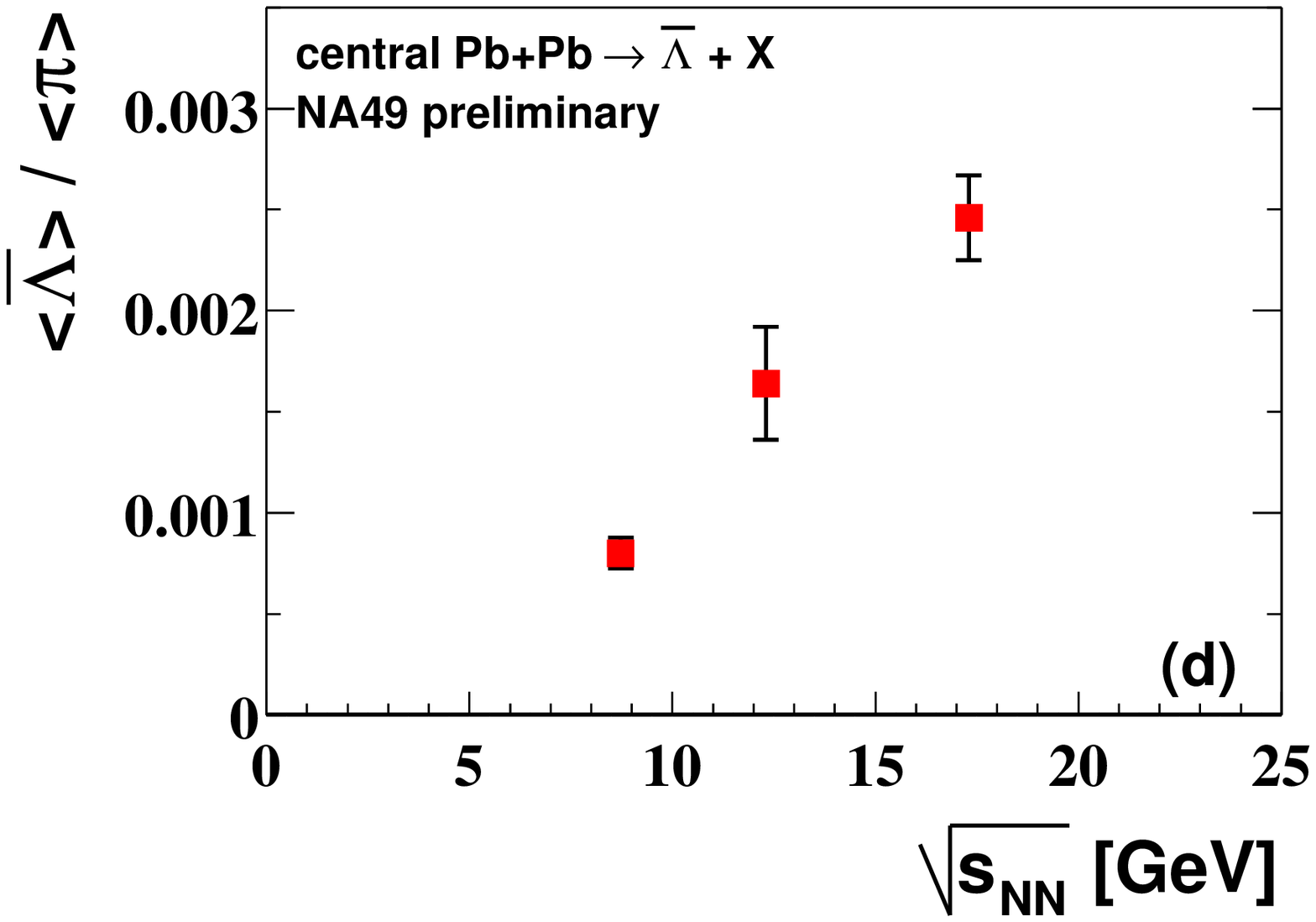}
 \end{center}
\end{minipage}
 \vspace{-1.2cm}
 \caption{\protect \footnotesize Energy dependence of the
$\Lambda/\pi$ (a) and the $\bar{\Lambda}/\pi$ (b) at mid-rapidity. 
In the logarithmic scaled inset of figure (b) the STAR measurement is
shown in addition. 
The same ratios are shown for the 4$\pi$ values (3(c) and (d)).  
The $\langle\Lambda\rangle / \langle\pi\rangle$ ratio from p+p
collisions at different energies~\cite{Gaz96,Gaz95} is shown as well. 
The different lines represent different model predictions.}
 \label{fig3}
\end{figure}
The maximum of the $\Lambda$ rapidity distribution decreases with
increasing collision energy, as shown in figure~\ref{fig1}. This
effect is even enhanced in the $\Lambda/\pi$ ratio (see
figure~\ref{fig3}(a)). The pions are calculated according to
$\pi = 3/2~(\pi^{+}+\pi^{-})$~\cite{Kla01,KPI02}.  
The data at AGS energies (triangles) are taken from
reference~\cite{Rai299,Alb02,Bac01,Pin02,Bec01}.
The $\Lambda/\pi$ ratio steeply increases at AGS energies, reaches a
maximum and drops at SPS energies. 
In comparison, the $\bar{\Lambda}/\pi$ ratio shows a monotonic
increase up to RHIC energies~\cite{STAR} without significant structure
(see figure~\ref{fig3}(b)). 
The same behavior is visible for the total multiplicities
$\langle\Lambda\rangle / \langle\pi\rangle$ and 
$\langle{\bar \Lambda}\rangle / \langle\pi\rangle$ as shown 
in figure~\ref{fig3}(c) and (d).
These differences can be attributed to different production mechanism
for $\Lambda$ and $\bar{\Lambda}$ and to the effect of net-baryon
density. 

Since the ${\rm K}^{+}$ carry about 80$\%$ of the produced 
$\bar{\rm s}$ 
quarks we expect the ${\rm K}^{+}/\pi^{+}$ ratio to show a
similar behavior as the $\Lambda/\pi$ ratio (using strangeness
conservation) which is indeed the case~\cite{KPI02}.  

In figure~\ref{fig3}, the measurements are compared to model
predictions from UrQMD~\cite{Web102}, HSD~\cite{Brat02},
RQMD~\cite{Xu02} and the statistical model of reference~\cite{pbm02}. 
All models describe the general trend of the experimental data
correctly. The microscopic models under-predict and the statistical
model over-predicts the measured $\Lambda/\pi$ ratios.\\

\begin{figure}[t]
\begin{minipage}[b]{6cm}
 \begin{center}
 \includegraphics[width=8.5cm,height=6.2cm]{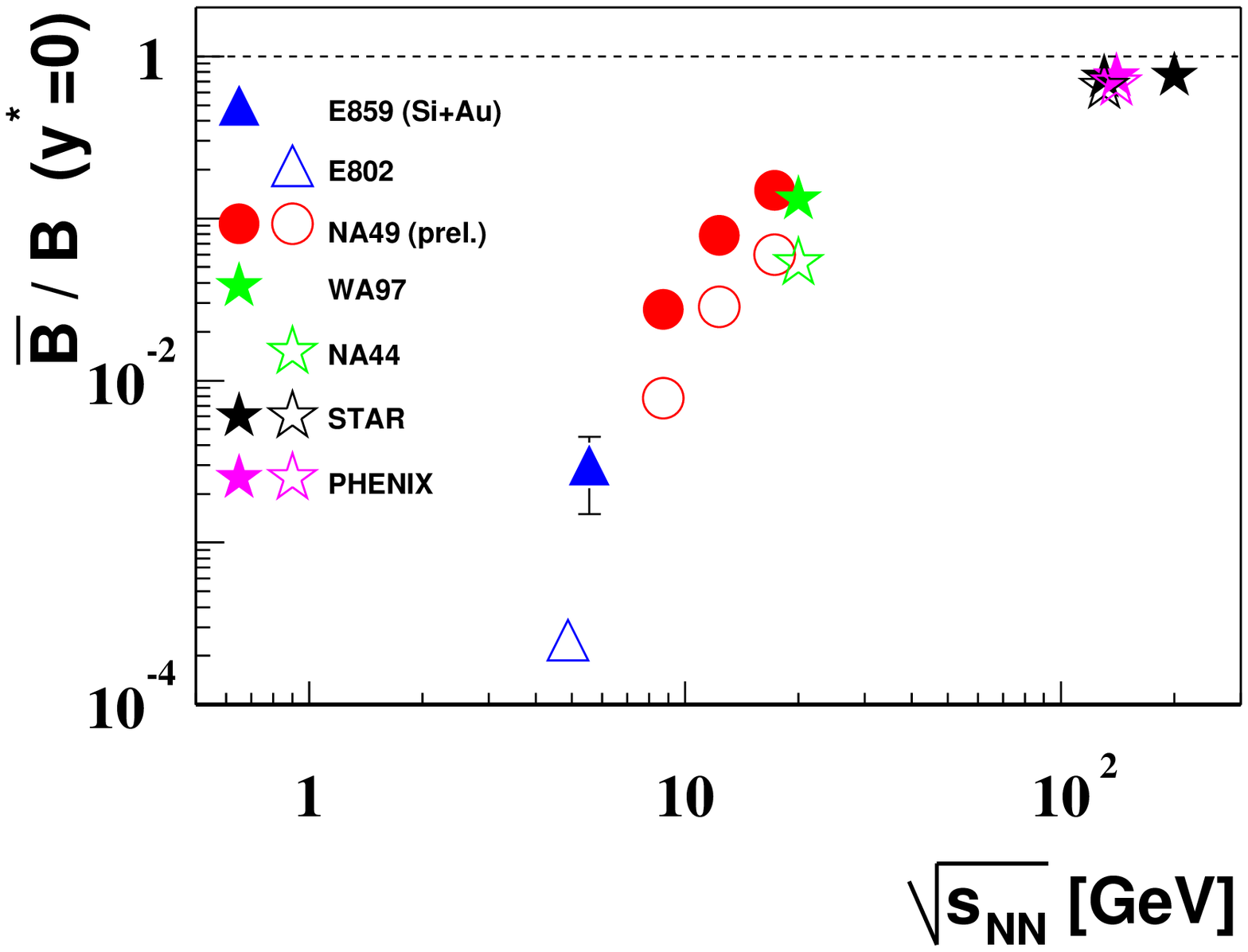}
 \end{center}
\end{minipage}
\hspace{2.cm}
%\hspace{\fill}
%
\begin{minipage}[b]{6cm}
 \begin{center}
 \includegraphics[width=8.cm,height=6cm]{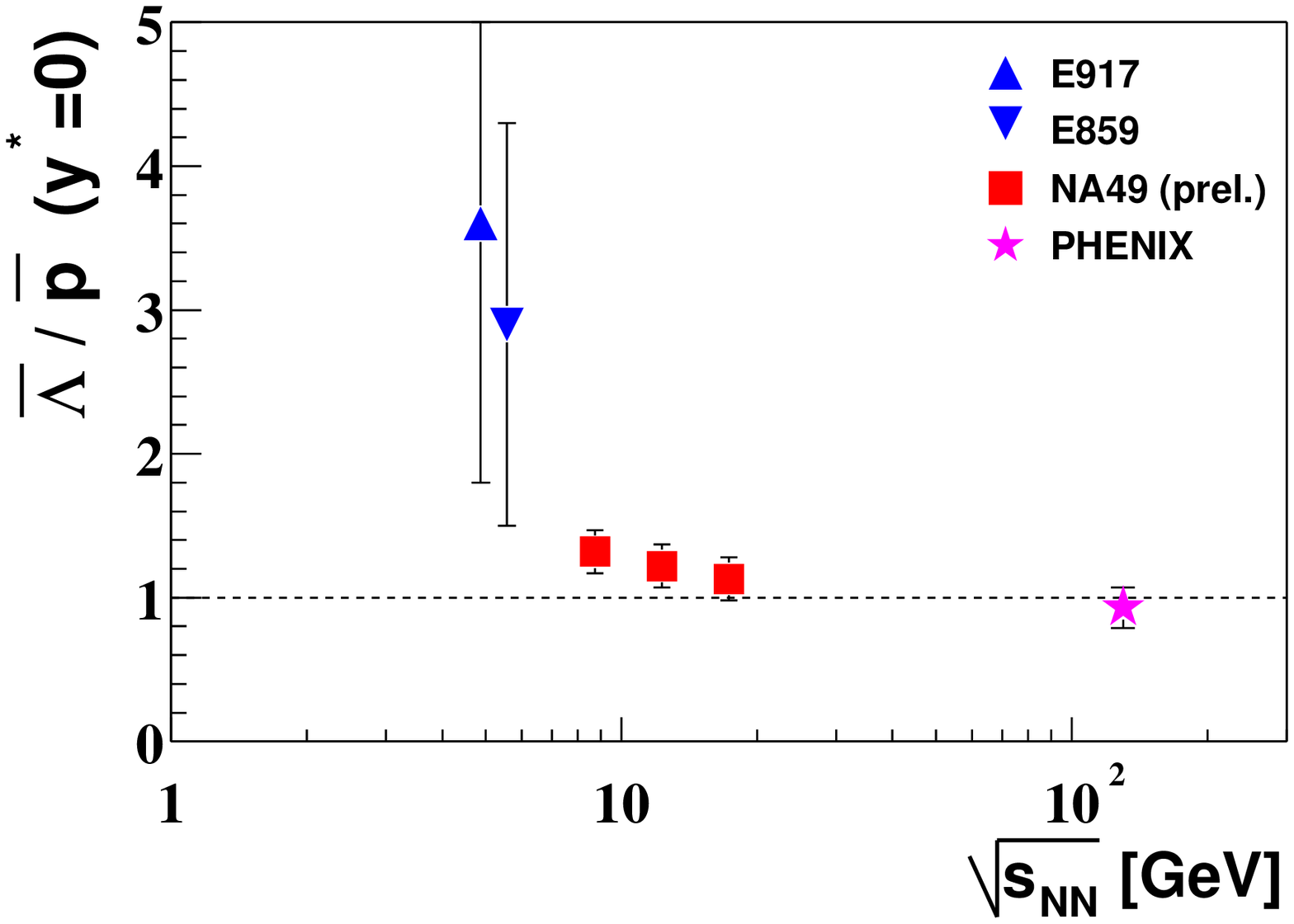}
 \end{center}
 \end{minipage}
 \vspace{-1.cm}
 \caption{\protect \footnotesize The $\bar{\Lambda}/\Lambda$ (left)
and $\bar{\Lambda}/\bar{\rm p}$ (right) ratio as a function of
c.m. energy. All particles are corrected for weak decay feed-down.}
 \label{fig5}
\end{figure}

The $\bar{\Lambda}/\Lambda$ ratio at mid-rapidity rises steeply 
from AGS~\cite{Aki95} to RHIC energies~\cite{STAR,PHEN02,Bur01} 
(see figure~\ref{fig5}, left). 
The numerical values are 
0.027 $\pm$ 0.0025 for 40, 
0.079 $\pm$ 0.01 for 80 and 
0.149 $\pm$ 0.016 for 158 A$\cdot$GeV, respectively.
The same trend is measured for the $\bar{\rm p}/{\rm p}$
ratio~\cite{Ahl98,Bea99}, but the numerical values are smaller than
for the corresponding $\bar{\Lambda}/\Lambda$ ratio.
(0.0079 $\pm$ 0.0008,  
0.028 $\pm$ 0.0025,
0.060 $\pm$ 0.005, respectively). 
Total yields for p and $\bar{\rm p}$ will give more insight into this
problem. 

The $\bar{\Lambda}/\bar{\rm p}$ ratio allows to study the interplay of
production and annihilation processes. In the SPS energy range this
ratio indicates an increase with decreasing energy as illustrated in
figure~\ref{fig5}, right. The data at lower and higher energies are
taken from~\cite{Bac01,Ste97,PHEN02}.\\ 

In summary, the ratio $\Lambda/\pi$ shows a maximum between top AGS
energies and 40~A$\cdot$GeV whereas the $\bar{\Lambda}/\pi$ ratio
increases monotonically with increasing energy. 
The upcoming measurements at 20 and 30~A$\cdot$GeV will give more
details for this energy range.    

%==============================================================================
\vspace{0.5cm}
\begin{footnotesize}
Acknowledgements:
The author thanks E.L. Bratkovskaya, K. Redlich and N. Xu 
for providing the model predictions.
This work was supported by the Director, Office of Energy Research,
Division of Nuclear Physics of the Office of High Energy and Nuclear Physics
of the US Department of Energy (DE-ACO3-76SFOOO98 and DE-FG02-91ER40609),
the US National Science Foundation,
the Bundesministerium fur Bildung und Forschung, Germany,
the Alexander von Humboldt Foundation,
the UK Engineering and Physical Sciences Research Council,
the Polish State Committee for Scientific Research (2 P03B 130 23 and 2 P03B 02418),
the Hungarian Scientific Research Foundation (T14920 and T32293),
Hungarian National Science Foundation, OTKA, (F034707),
the EC Marie Curie Foundation,
and the Polish-German Foundation.    
\end{footnotesize}

%==============================================================================
\begin{footnotesize}
%\bibliography{literatur.bib}

\end{footnotesize}

%==============================================================================
\end{document}